\documentclass[pdftex,twocolumn,epjc3]{svjour3}

\usepackage{graphicx}
\usepackage{mathptmx,amsmath,amsfonts,amssymb}
\usepackage[T1]{fontenc}
\usepackage{flushend}
\usepackage[numbers,sort&compress]{natbib}
\usepackage[colorlinks,citecolor=blue,urlcolor=blue,linkcolor=blue]{hyperref}

\journalname{Eur. Phys. J. C}

\begin{document}
	
\title{ Spherical and cylindrical solutions in $f(T)$ gravity by Noether Symmetry Approach}

\author{Ali Nur Nurbaki\thanksref{e1,e2,addr1,addr6}
        \and
        Salvatore Capozziello\thanksref{e3,addr2,addr3,addr4}
        \and
        Cemsinan Deliduman\thanksref{e4,addr5} }

\thankstext{e1}{e-mail: ali.nurbaki@ogr.istanbul.edu.tr}

\thankstext{e2}{e-mail: anurbaki@yahoo.com}

\thankstext{e3}{e-mail: capozziello@na.infn.it}

\thankstext{e4}{e-mail:  cemsinan@msgsu.edu.tr}

\institute{Dept. of Astoronomy \& Space Sciences, Istanbul University, 34134 Beyazit, Istanbul, Turkey\label{addr1}
          \and
          Dipartimento di Fisica  ``E. Pancini", Universit\`a di Napoli  ``Federico II", Via Cinthia, I-80126, Napoli, Italy,\label{addr2}
          \and
Istituto Nazionale di Fisica Nucleare (INFN), sez. di Napoli, Via Cinthia 9, I-80126 Napoli, Italy,\label{addr3}
          \and
Laboratory for Theoretical Cosmology, Tomsk State University of Control Systems and Radioelectronics (TUSUR), 634050 Tomsk, Russia,\label{addr4}
        \and
          Department of Physics, Mimar Sinan Fine Arts University, Bomonti, Istanbul 34380, Turkey.\label{addr5}
       \and
       Institute of Graduate Studies in Sciences, Istanbul University, 34134, Beyazit, Turkey \label{addr6} }
    
\date{Received: date / Accepted: date}

\maketitle

\abstract
We find exact  solutions for  $f(T)$  teleparallel gravity for the cases of spherically and cylindrically symmetric tetrads. The adopted method is based on the search for  Noether symmetries of point-like Lagrangians defined in  Jordan and  Einstein frames. Constants of motion  are used to reduce the dynamical system.  We first consider the Lagrangian defined in the  Jordan frame  for a spherically symmetric tetrad and, by the help of two constants of motion,  we eliminate a tetrad potential and integrate the other. The more complicated structure in the Einstein frame is also overcome by the same method. After that we obtain the Jordan frame Lagrangian for a general cylindrically symmetric tetrad. Following the same procedure adopted in the  spherically symmetric case,  we again obtain the tetrad potentials and then the exact solutions.



\section{Introduction}

Although General Relativity is accepted to be the most successful theory of gravity, due, for example, to the recent discovery of gravitational waves confirming the validity of the Einstein approach, it is worth noticing  that this success is valid observationally below the solar system scale and theoretically far away from the Planck scale. We can roughly say that General Relativity has some observational shortcomings at infrared (IR) and theoretical shortcomings at ultraviolet (UV) scale. For example, at the UV scale, early universe cosmology with General Relativity necessitates an early inflation era in order to determine conditions for today's universe. Other than that, renormalization of gravity at the quantum regime and unification with other forces requires the General Relativity has to be modified. At the IR scale, on the other hand, the theoretical  prediction of accelerated expansion of the Universe with General Relativity requires either the help of a source term, the so called "dark energy," or a cosmological constant term. Similarly, the case of galactic rotation dynamics or the dynamics of the galactic clusters analyzed with General Relativity again calls for the help of source terms, namely the "dark matter". Despite the evidence of these dark ingredients at astrophysical and cosmological scales, they continue to escape any finding at fundamental scales.

Many researchers now look for a source for dark components not in the matter sector, but in geometry by taking into account modified theories of gravitation. There have been many attempts to extend Einstein gravity to  the UV and the IR scales \cite{report}. However, how the extension is done depends very much on the motivation.   For example one can extend General Relativity to five dimensions in order to unify it with Electromagnetism. Another way of extension is to add scalar field(s) to the Lagrangian as it is done in Brans--Dicke theory to obtain a fully Machian theory. Last but not least method of extending General Relativity is adding higher order geometric invariants to the Lagrangian. This was the case when first attempts were performed to renormalize General Relativity. Then, independent of renormalization concerns, mostly for observational motivations, higher-order terms were introduced as general functional forms $f(R)$ in the Lagrangian. This  theories are called Extended Theories of Gravity if General Relativity belongs to the group or Modified Theories of Gravity if the Einstein approach is not retained.

All these alternative  theories of gravity can be considered   beyond the standard General Relativity when conflicts with observations are taken into account. In a naive approach, these conflicts with  observations can be  palliatively solved by adding dark matter and dark energy terms to the source sector of Einstein's equations. On the other hand, Extended  Theories of Gravity  deal with  dark matter and dark energy as curvature effects and define modifications of the geometric sector of the equations \cite{Capozziello:2006uv,Capozziello:2012ie,DelidumanDM,Nojiri:2006ri}. What is used is not always the functional form of the Ricci curvature scalar  $R$; general functions of different geometric invariants as the torsion scalar $T$, the Gauss-Bonnet topological invariant $\cal G$, the Weyl scalars,   etc. are also in the class of modified gravity theories. 

Here we  take into account   a theory of gravitation with torsion, the so called "Teleparallel Equivalent of General Relativity (TEGR)", \cite{pereira} and its extension, the so-called  $f(T)$  gravity. The aim is to obtain exact solutions adopting the Noether Symmetry Approach \cite{Capozziello:1996bi}. In particular, we are interested in obtaining spherically and cylindrically symmetric solutions.

The layout of the paper is the following.  The main ingredient of TEGR are summarized in Sec. 2.
Sec. 3 is devoted to recast $f(T)$ gravity in scalar-tensor form in order to deal with conformal transformations. 
The Noether Symmetry Approach for our scalar-tensor form of $f(T)$ is discussed in Sec.4. Spherical solutions in Jordan and Einstein frames are derived in Sec. 5. Solutions for a cylindrical symmetric tetrad are derived in Sec. 6. Conclusions are drawn in Sec. 7.


\section{Teleparallel Equivalent General Relativity  and its $f(T)$ extension }

After  $f(R)$ gravity gained attention in the context of modified gravities, it has been realized   that modification of gravity cannot be  limited only to functions of Ricci scalar  curvature invariant $R$. In fact, also  functional forms of other geometric invariants such as torsion or Gauss-Bonnet invariants can be useful to address the problem of gravity at UV and IR scales \cite{ferInfl,Nojiri:fRTG,Nojiri:2007frg,NoetherG,FluidG}. Recently one of these theories, namely the $f(T)$ gravity,  gained much attention amongst the others because it was shown to explain cosmological phenomena such as the early inflation [\citealp{ferInfl}] and the late acceleration [\citealp{linDE}] of the universe (for a review see [\citealp{Cai}]). Besides, there is the possibility to explain galactic dark matter phenomena via $f(T)$ gravity [\citealp{rahDM,Finch}].

Gravitational theories with torsion were developed first by Einstein himself. Being a gauge theory of gravity and giving an equivalent physics to General Relativity, it is called "Teleparallel Equivalent of General Relativity". The dynamical variable of the theory is the tetrad field and the connection is the Weitzenb\"{o}ck connection
\begin{equation}
\label{we} 
\Gamma^{\rho}_{\alpha\beta} \equiv
e^{\rho}_{A}\partial_{\alpha}e^{A}_{\beta},
\end{equation}
which gives zero curvature but non-zero torsion. Torsion tensor is
\begin{equation}
\label{wt} 
T^{\rho}_{\alpha\beta}=\Gamma^{\rho}_{\alpha\beta}-\Gamma^{\rho}_{\beta\alpha}\equiv e^{\rho}_{A}[\partial_{\alpha}e^{A}_{\beta}-\partial_{\beta}e^{A}_{\alpha}].
\end{equation}
Torsion scalar can be computed such that
\begin{equation}
\label{t4} T \equiv {S_{\rho}}^{\alpha\beta} {T^{\rho}}_{\alpha\beta},
\end{equation}
where
\begin{equation}
\label{t5} {S_{\rho}}^{\alpha\beta} \equiv  \frac{1}{2}[{K^{\alpha\beta}}_{\rho}+{\delta}^{\alpha}_{\rho}{T^{\gamma\beta}}_ {\gamma}-{\delta}^{\beta}_{\rho}{T^{\gamma\alpha}}_{\gamma}],
\end{equation}
and
\begin{equation}
\label{t3} K^{\alpha\beta}_{ \rho} \equiv  -\frac{1}{2}[{T^{\alpha\beta}}_{\rho}-{T^{\beta\alpha}}_ {\rho}-{T_{\rho}}^{\alpha\beta}].
\end{equation}
\noindent
Gravitational Lagrangian in TEGR is represented by the torsion scalar $T$ and thus the action is
\begin{equation}
\label{Tint} {I} = \int  d^4 x   e  T .
\end{equation}
Like in $f(R)$ theory of gravity, action \eqref{Tint} can be straightforwardly generalized to 
\begin{equation}
\label{t6} {I} = \int  d^4 x   e   f(T)+ I_m ,
\end{equation}
where $e$ = det $e^{i}_{\mu}=\sqrt {-g}$, $I_m$ is the matter action, and the units can be  chosen so that $c = 16 \pi G = 1$. This Lagrangian gives us the so-called $f(T)$ theory of gravity. It has first been proposed to explain the  early inflation of the Universe, and then it is observed that this theory could explain the late time cosmic acceleration and the dark matter phenomena. For the other physical motivations, the reader can see the book \cite{pereira}.


\section{$f(T)$ theory in scalar--tensor form}

It is possible to recast  $f(T)$ Lagrangian in an equivalent scalar-tensor form by choosing a suitable scalar field \cite{Yang}.
A general form of the action \eqref{t6} can be written as
\begin{equation} \label{f1}
I= \int d^4x e [f(\chi)+(T-\chi)f'(\chi)]+I_{\rm m}(e^i_\mu),
\end{equation}
where $f'(\chi)\equiv df/d\chi$. It is straightforward to see that if $f''(\chi)\neq 0$ then the field equation for $\chi$, i.e. $(T-\chi)f''(\chi)=0$, is solved by $\chi=T$. Substitution of this into (\ref{f1}) gives back the action (\ref{t6}). If we introduce a scalar field $\varphi=\varphi(r)$ satisfying $F(\varphi)=f'(\chi)$, then (\ref{f1}) can be written in the form of an action for a scalar--tensor theory:
\begin{equation} \label{f2}
I_{\rm JF}=\int d^4x e [F(\varphi)T-\omega(\varphi)g^{\mu\nu}\nabla_\mu \varphi \nabla_\nu \varphi-V(\varphi)]
+I_m (e^{i}_{\mu}),
\end{equation}
where $V(\varphi)=\chi f'(\chi)-f(\chi)$ is the scalar potential. 
This form of the action is in  the Jordan frame since it contains non-minimally coupled terms. 

However, an important remark is in order at this point. As discussed in details in \cite{Cai},  $f(T)$ and $f(R)$ gravity are deeply different. For example, $f(R)$gravity gives rise to fourth-order field equations while $f(T)$ remains of second order like standard TEGR with  $f(T)= T$. Furthermore,   one cannot univocally write
$f(T)$ gravity as a torsion scalar and a scalar field, as in the case of  $f(R)$ which can be always  recast as a curvature scalar and a scalar field. This is  because $f(T)$ is not invariant under local Lorentz transformation.  This feature of the theory means that one can generate a torsion scalar $T$ with a scalar field and, at the same time, one can generate
another $\tilde{T}$, for the same spacetime,  with another scalar field.  In other words, the scalar-tensor form of $f(T)$ gravity assumes physical meaning only if a specific tetrad is defined and, starting from this, a related metric is derived as we will do below. Thanks to these considerations, it is realistic to search for solutions in $f(T)$ gravity.

Applying a conformal transformation, one can obtain an equivalent action which equals to Einstein--Hilbert term plus a scalar field, i.e., the action of a scalar--tensor theory in the Einstein frame. In some works \cite{ConformalWright,Yang},  it is investigated whether $f(T)$ gravity behaves in the same fashion as $f(R)$ gravity  under a conformal transformation. Furthermore, as pointed out in \cite{Bamba}, a conformal formulation  can always be adopted  in a pure and extended TEGR, that is for an action containing just $T$ or its extension $f(T)$. Specifically, these authors  propose conformal scalar and gauge theories in TEGR and study the conformal torsion gravity. They  demonstrate the existence  of  cosmological solutions like power-law acceleration and de Sitter expansion, realized in the framework of $f(T)$ gravity where a  conformal scalar field  and  a conformal torsion  are present.

Following these works, it is possible to write  the conformal transformation for the torsion scalar as
\begin{equation} \label{eq1}
\hat{T}=\Omega^{-2}T+4\Omega^{-3}\partial^{\mu}\Omega T^{\rho}_{\phantom{\rho}\rho\mu}-6\Omega^{-4}\partial_{\mu}\Omega\partial^{\mu}\Omega
\end{equation}
and then
\begin{equation} \label{eq2}
T=\Omega^{2}\hat{T}-4\Omega\partial^{\mu}\Omega\hat{T}^{\rho}_{\phantom{\rho}\rho\mu}-6\hat{\partial}^{\mu}\Omega\hat{\partial}_{\mu}\Omega,
\end{equation}
where we used 
\begin{equation} \label{cont1}
\hat{g}_{\mu\nu}=\Omega^2(x)g_{\mu\nu},\quad \hat{g}^{\mu\nu}=\Omega^{-2}(x)g^{\mu\nu}, 
\end{equation}
and
\begin{equation} \label{cont2}
\hat{e}^{a}_\mu=\Omega(x)e^{a}_\mu,\quad \hat{e}^{\mu}_a=\Omega^{-1}(x)e^{\mu}_a,\quad \hat{e}=\Omega^4e .
\end{equation}
By using ~\eqref{eq2} in ~\eqref{f2} we obtain a scalar--tensor theory in the Einstein frame as

\begin{eqnarray} \label{Lageins}
I_{EF} &=& \int d^{4}x~e\left[ \hat{T}-2\frac{[F^{'}(\varphi)]}{F}\hat{\nabla}^{\mu}\varphi\hat{T}^{\rho}_{\phantom{\rho}\rho\mu} \right. \\
&& \left. -(\frac{2\omega}{F}+\frac{3[F^{'}(\varphi)]^{2}}{F^{2}})\frac{1}{2}\hat{g}^{\mu\nu}\hat{\nabla}_{\mu}\varphi\hat{\nabla}_{\nu}\varphi-U(\varphi) \right]+I_{m} ,\nonumber
\end{eqnarray}\\
or alternatively

\begin{eqnarray} \label{altLag}
I_{EF } &=& \int d^{4}x~e \left[ \hat{T}+2F^{-1}\hat{\partial}^{\mu}F~\hat{T}^{\rho}_{\phantom{\rho}\rho\mu}  -\frac{1}{2}\hat{g}^{\mu\nu}\hat{\nabla}_{\mu}\psi\hat{\nabla}_{\nu}\psi-U(\varphi) \right] \nonumber \\
&& +I_{m}[F(\varphi)^{-1/2}\hat{e}^{i}_{\phantom{i}\mu} ] ,
\end{eqnarray}
where 
\[
F=\Omega^{2},\quad U=\frac{V(\varphi)}{F(\varphi)^{2}} ,
\] 
and 
\[
\left(\frac{d\psi}{d\varphi}\right)^{2}=\left(\frac{2\omega}{F}+\frac{3[F^{'}(\varphi)]^{2}}{F^{2}}\right) .
\]
In Ref.\cite{Yang},  the Lagrangian in the Einstein frame  is different from (\ref{altLag}). If we vary matter free form of this  Lagrangian (where $I_{m}$ vanishes) with respect to $\hat{e}^{i}_{\phantom{i}\lambda}$,  we obtain the following field equations 

\begin{eqnarray} \label{eq13}
\hat{\partial}_{\mu}(\hat{e}S_{i}^{\phantom{i}\mu\lambda})
+\hat{e}\hat{e}_{i}^{\phantom{i}\nu}\hat{T}^{\rho}_{\phantom{\rho}\mu\nu}\hat{S}_{\rho}^{\phantom{\rho}\mu\lambda}
-\frac{1}{4}\hat{e}\hat{e}_{i}^{\phantom{i}\lambda}\hat{T} \nonumber && \\ +\frac{1}{2}\partial_{\mu}(\hat{e}F^{-1}\hat{\partial}^{\mu}F\hat{e}_{i}^{\phantom{i}\lambda}-\hat{e}F^{-1}\hat{\partial}^{\lambda}F\hat{e}_{i}^{\phantom{i}\mu}) && \nonumber \\ +\frac{1}{2}\hat{e}\hat{e}_{i}^{\phantom{i}\lambda}F^{-1}\hat{\partial}^{\mu}\hat{T}^{\rho}_{\phantom{\rho}\rho\mu}
+\frac{1}{2}\hat{e}\hat{e}_{i}^{\phantom{i}\rho}F^{-1}\hat{\partial}^{\mu}\hat{T}^{\lambda}_{\phantom{\lambda}\rho\mu} && \nonumber \\ +\frac{1}{4}\hat{e}\hat{e}_{i}^{\phantom{i}\nu}\hat{\nabla}^{\lambda}\psi\hat{\nabla}_{\nu}\psi 
-\frac{1}{8}\hat{e}\hat{e}_{i}^{\phantom{i}\lambda}\hat{\nabla}^{\mu}\psi\hat{\nabla}_{\mu}\psi && \nonumber \\ -\frac{1}{4}\hat{e}\hat{e}_{i}^{\phantom{i}\lambda}U(\psi)&=&0 .
\end{eqnarray}\\

By a rapid inspection  of the field equations, we see that the first three terms come from the torsion scalar $T$ and the last three terms are related with the scalar field. These six terms are the expected terms for a scalar--tensor theory in the Einstein frame. But the remaining three terms, coming from the torsion tensor,  prevents us from obtaining true Einstein frame field equations.

Using the first form of the Einstein frame action (\ref{Lageins}) and by defining a spherically symmetric tetrad as
\begin{equation} \label{tet1}
	{e}^{i}_{\phantom{i}\nu}=
	\begin{bmatrix}
	A(r) & 0 & 0 & 0\\
	0 & B(r) & 0 & 0\\
	0 & 0 & M(r) & 0\\
	0 & 0 & 0 & M(r) \sin \theta
	\end{bmatrix} ,
\end{equation}
we obtain the point-like Lagrangian in the Einstein frame given as
\begin{eqnarray} \label{LagEin}
	L_{EF} = \sin\theta ( &-& A B M^2 U +\frac{\varphi_r^2 A M^2 \omega(\varphi)}{F B}  \nonumber \\
	&+& \frac{2 M_r^2 A }{B}-\frac{4 A_r M_r M}{B}+\frac{4 \varphi_r M_r A B M }{F(\varphi)} \nonumber \\
	&+&  \frac{2 \varphi_r A_r F_\varphi B M^2 }{F(\varphi)}-\frac{3\varphi_r^2 F_\varphi^2 A M^2}{F^2 B} ) .
\end{eqnarray}
Where lower index indicates differentiation. Similar to \eqref{LagEin} a point-like Lagrangian for the Jordan frame can be obtained from \eqref{f2} and \eqref{tet1} as

\begin{eqnarray} \label{LagJor}
	L_{JF} &=& -\frac{\varphi_r^2 A M^2 \omega(\varphi) \sin\theta}{B}-\frac{2 M_r^2 A F(\varphi)\sin\theta}{B} \nonumber \\
	&& -\frac{4 A_r M_r M F(\varphi) \sin\theta}{B}-A B M^2 V(\varphi)\sin\theta ,
\end{eqnarray}
which is canonical but singular. The singularity emerges due to the fact that this Lagrangian does not contain any term with $B_r$.


\section{The Noether Symmetry Approach}

Noether symmetries have become a standard tool in mathematical physics. They are useful both for simplification of differential equations and determination of integrable systems. In our context, the Noether Symmetry Approach  is often used in modified and extended gravity theories for constraining the functional form of the Lagrangian densities $f(R), f(T)$, $f(R,{\cal G})$ etc., as well as finding solutions of the field equations from such point-like Lagrangians \cite{capAxial,Capozziello:1996bi,Basilakos:2013rua,Bahamonde1,Bahamonde2,Bahamonde3,Kostas}.

Noether symmetry exists for a Lagrangian $L$ such that Lie derivative along a Noether vector $X$ vanishes for $L$:
\begin{equation}
\mathcal{L}_{X}{L}=0,
\end{equation}
namely 
\begin{equation}
X {L}=0\,.
\end{equation}
In our case, we define the Noether vector as a general infinitesimal symmetry generator given by
\begin{equation} \label{noe}
 X = \alpha\partial_{A}+\beta \partial_{B}+\gamma\partial_{\varphi}+\delta \partial_{M} 
 +\dot{\alpha}\partial_{\dot{A}}+\dot{\beta} \partial_{\dot{B}}+\dot{\gamma}\partial_{\dot{\varphi}}+\dot{\delta}\partial_{\dot{M}},
 \end{equation}
 where overdot indicates differentiation with respect to an affine parameter (in this case $r$) and
 
 \begin{eqnarray}
 \dot{\alpha}&=\Big(\frac{\partial \alpha}{\partial A}\Big)\dot{A}+\Big(\frac{\partial \alpha}{\partial B}\Big)\dot{B}+\Big(\frac{\partial \alpha}{\partial \varphi}\Big)\dot{\varphi}+
 \Big(\frac{\partial \alpha}{\partial M}\Big)\dot{M}\,,\\
 \dot{\beta}&=\Big(\frac{\partial \beta}{\partial A}\Big)\dot{A}+\Big(\frac{\partial \beta}{\partial B}\Big)\dot{B}+\Big(\frac{\partial \beta}{\partial \varphi}\Big)\dot{\varphi}+
 \Big(\frac{\partial \beta}{\partial M}\Big)\dot{M}\,,\\
 \dot{\gamma}&=\Big(\frac{\partial \gamma}{\partial A}\Big)\dot{A}+\Big(\frac{\partial \gamma}{\partial B}\Big)\dot{B}+\Big(\frac{\partial \gamma}{\partial \varphi}\Big)\dot{\varphi}+
 \Big(\frac{\partial \gamma}{\partial M}\Big)\dot{M}\,,\\
 \dot{\delta}&=\Big(\frac{\partial \delta}{\partial A}\Big)\dot{A}+\Big(\frac{\partial \delta}{\partial B}\Big)\dot{B}+\Big(\frac{\partial \delta}{\partial \varphi}\Big)\dot{\varphi}+
 \Big(\frac{\partial \delta}{\partial M}\Big)\dot{M}\,.
 \end{eqnarray}
 \\
 A Noether symmetry leads to the existence of a constant of motion given by
\begin{equation} 
\label{cons}
\Sigma = \alpha\frac{\partial L }{\partial\dot A}+ \beta\frac{\partial L }{\partial\dot B}+ \gamma \frac{\partial L }{\partial\dot \varphi}+ \delta\frac{\partial L }{\partial\dot M} .
\end{equation}
We choose to represent $\alpha, \beta, \gamma, \delta$ functions as components of a vector defined by
\begin{equation}
\label{nvec}
\stackrel{\to }{N}=(\alpha, \beta, \gamma, \delta)\, ,
\end{equation}
which corresponds to the generating vector \eqref{noe}. If we find a set of components for $\stackrel{\to }{N}$ by using $X {L}=0$, then we can find a way to integrate the functions $A, B, M, \varphi$ by constructing constants of motion. 
In the next section,  we are going to apply Noether symmetries to the Lagrangians defined in both Jordan and Einstein frames for a general spherically symmetric tetrad.


\section{Spherical solutions in Jordan and Einstein frame  via Noether symmetries}

\subsection{Jordan frame}

By using the point-like Lagrangian for the Jordan frame (\ref{LagJor}) we write down the Noether equation, $X L = 0$, equating the coefficients of $\dot{A}^2$, $\dot{B}^2$, $\dot{M}^2$, $\dot{\varphi}^2$ and the cross terms like $\dot A \dot B$, $\dot A \dot M$, $\dot A \dot \varphi$, $\dot B \dot M$, $\dot B \dot \varphi$, $\dot M \dot \varphi$ etc. along with the term free from derivative with respect to $r$ to zero. Doing so we obtained a set of 11 partial differential equations for $F(\varphi),$ $\omega(\varphi)$, and $V(\varphi).$ Not all these equations are independent however. Independent equations are
\begin{equation}\label{ff1} 
\delta_A=0,\\ \delta_B=0,\\ \alpha_B=0,\\ \gamma_B=0 ,
\end{equation}

\begin{equation}\label{ff2} 
\frac{F_{\varphi}}{F}\gamma+\delta_{M}+\alpha_{A}-\frac{\beta}{B}+\frac{\delta}{M}+\frac{A}{M}\delta_A=0 ,
\end{equation}

\begin{equation}\label{ff3} 
\frac{\omega_{\varphi}}{\omega}\gamma+\gamma_\varphi+2\frac{\delta}{M}+\frac{\alpha}{A}-\frac{\beta}{B}=0 ,
\end{equation}

\begin{equation}\label{ff6}
	\frac{\omega}{F}AM\gamma_A+2\delta_\varphi=0 ,
\end{equation}

\begin{equation}
	\frac{\omega}{F}M\gamma_M+2\frac{\alpha_\varphi}{A}+2\frac{\delta_\varphi}{M}=0 ,
\end{equation}

\begin{equation}\label{ff4} 
\frac{V_{\varphi}}{V}\gamma+2\frac{\delta}{M}+\frac{\alpha}{A}+\frac{\beta}{B}=0 ,
\end{equation}

\begin{equation}\label{ff5}
\frac{F_{\varphi}}{F}\gamma+2\delta_{M}-\frac{\beta}{B}+\frac{\alpha_{A}}{A}+2\frac{M}{A}\alpha_M=0.
\end{equation}

We can find two linearly independent vectors satisfying these equations, which are
\begin{equation}
\label{noevec}
\stackrel{\to }{N_1}=(A, B, \varphi, M)\,,\\
\stackrel{\to }{N_2}=(3A, B, \varphi, M).
\end{equation}
Thus, $F(\varphi), \omega(\varphi)$, and $V(\varphi)$ can be solved from equations \eqref{ff1}-\eqref{ff5} for these Noether vectors as
\begin{equation}
\label{fvom}
F(\varphi)=\frac{c_1}{\varphi^2},\\ \omega(\varphi)=\frac{c_2}{\varphi^3},\\ V(\varphi)=\frac{c_3}{\varphi^4} .
\end{equation}
We know that for commuting Noether vectors of a system, we can construct multiple constants of motion for that system (see \cite{CapoFaraoniBeyond} for details).
By using \eqref{cons} and \eqref{noevec} we obtain two constants of motion, $\Sigma_1$ and $\Sigma_2$, given by
\begin{eqnarray} \label{sigma1}
	\Sigma_1 &=& -2\sin(\theta)\frac{\varphi\varphi_rAM^2\omega}{B}-8\sin(\theta)\frac{MM_rFA}{B} \nonumber \\
	&&-4\sin(\theta)\frac{A_rFM^2}{B} ,
\end{eqnarray}
\begin{equation}\label{sigma2}
	\Sigma_2=-2\sin(\theta)\frac{\varphi\varphi_rAM^2\omega}{B}-12\sin(\theta)\frac{MM_rFA}{B} .
\end{equation}
Following [\citealp{capAxial}], we can integrate $A_r$ to obtain $A$. For this,  we need to eliminate $B$ from \eqref{sigma1} and \eqref{sigma2}. 
By using (\ref{sigma2}) we obtain $B$ as
\begin{equation}
\label{BB}
	B=-2\sin(\theta)\frac{(\varphi\varphi_rAM^2\omega+6MM_rFA)}{\Sigma_2} .
\end{equation}
We can substitute this expression into \eqref{sigma1} together with \eqref{fvom} and, after defining $M=r$, we arrive to the expression
\begin{equation}
	\Sigma_1(c_2r\frac{\varphi_r}{\varphi^2}+6\frac{c_1}{\varphi^2})=\Sigma_2(c_2r\frac{\varphi_r}{\varphi^2}+4\frac{c_1}{\varphi^2}+2\frac{c_1r}{\varphi^2}\frac{A_r}{A}) .
\end{equation}
After integrating this equation,  we obtain $A(r)$ as
\begin{equation}
\label{AA}
	A (r)=\frac{e^\frac{(\varSigma_1-\varSigma_2)c_2\varphi}{2c_1}}{Kr^\frac{2\varSigma_2}{3\varSigma_1}} .
\end{equation}
In the case of $\varphi=ln(r)$, this expression turns out to be
\begin{equation}
\label{Ar}
A(r)=\frac1{K} r^ {\frac{(\varSigma_1-\varSigma_2)c_2}{2c_1}-\frac{2\varSigma_2}{3\varSigma_1}}  .
\end{equation}
By substituting (\ref{Ar}) into (\ref{BB}), we finally obtain $B(r)$ as
\begin{equation}
\label{Br}
B(r)=\frac{-2\sin(\theta)(c_2+6c_1)}{K\varSigma_2}\frac{r^ {1+\frac{(\varSigma_1-\varSigma_2)c_2}{2c_1}-\frac{2\varSigma_2}{3\varSigma_1}}}{ln(r)^2} .
\end{equation}
By obtaining the potentials $A(r)$, $B(r)$ via Noether symmetry method, we thus  find the spherically symmetric solution in the Jordan frame.


\subsection{Einstein Frame}

By using the point-like Lagrangian for the Einstein frame \eqref{LagEin} in the Noether equation, $X L = 0$, and equating the coefficients of the quadratic first order derivatives to zero as it is done in the Jordan frame, we obtain the following set of equations:

\begin{eqnarray} \label{Nein1}
	-(\frac{2AM^2\omega}{FB}+\frac{3F'^2AM^2}{F^2B})\gamma_M+\frac{4F'ABM}{F}\gamma_\varphi && \nonumber \\
	+(\frac{4F"ABM}{F}-\frac{4F'^2ABM}{F^2})\gamma+\frac{2F'BM^2}{F}\alpha_M && \nonumber \\
	+\frac{4F'ABM}{F}\delta_M+\frac{4F'BM}{F}\alpha-\frac{4M}{B}\alpha_\varphi && \nonumber \\
	+\frac{4F'AM}{F}\beta+\frac{4F'AB}{F}\delta-\frac{4A}{B}\delta_\varphi &=& 0 ,
\end{eqnarray}

\begin{eqnarray}  \label{Nein2}
	-(\frac{2AM^2\omega}{FB}+\frac{3F'^2AM^2}{F^2B})\gamma_A+\frac{2F'BM^2}{F}\gamma_\varphi && \nonumber \\
	+(\frac{2F"BM^2}{F}-\frac{2F'^2BM^2}{F^2})\gamma+\frac{2F'BM^2}{F}\alpha_A && \nonumber \\
	+\frac{4F'ABM}{F}\delta_A+\frac{2F'BM^2}{F}\beta && \nonumber \\
	+\frac{4F'BM}{F}\delta-\frac{4M}{B}\delta_\varphi &=& 0 ,
\end{eqnarray}

\begin{eqnarray} \label{Nein3}
	-(\frac{2AM^2\omega}{FB}+\frac{3F'^2AM^2}{F^2B})\gamma_\varphi-\frac{AM^2\omega_\varphi}{FB}\gamma && \nonumber \\
	+\frac{F'AM^2\omega}{F^2B}\gamma-\frac{3F'F"AM^2}{F^2B}\gamma+\frac{3F'^3AM^2}{F^3B}\gamma && \nonumber \\
	-(\frac{M^2\omega}{FB}+\frac{3F'^2M^2}{2F^2B})\alpha+\frac{2F'BM^2}{F}\alpha_\varphi && \nonumber \\
	+(\frac{AM^2\omega}{FB^2}+\frac{3F'^2AM^2}{2F^2B^2})\beta && \nonumber \\
	-(\frac{2AM\omega}{FB}+\frac{3F'^2AM}{F^2B})\delta+\frac{4F'ABM}{F}\delta_\varphi &=& 0 ,
\end{eqnarray}

\begin{equation} \label{Nein4}
	\frac{4F'ABM}{F}\gamma_M-\frac{4M}{B}\alpha_M-\frac{4A}{B}\delta_M-\frac{2}{B}\alpha+\frac{2A}{B^2}\beta=0 ,
\end{equation}

\begin{eqnarray} \label{Nein5}
	\frac{2F'BM^2}{F}\gamma_M+\frac{4F'ABM}{F}\gamma_A-\frac{4M}{B}\alpha_A+\frac{4M}{B^2}\beta && \nonumber \\
	-\frac{4A}{B}\delta_A-\frac{4}{B}\delta-\frac{4M}{B}\delta_M&=& 0 ,
\end{eqnarray}

\begin{eqnarray} \label{Nein6}
-(\frac{2AM^2\omega}{FB}+\frac{3F'^2AM^2}{F^2B})\gamma_B+\frac{2F'BM^2}{F}\alpha_B && \nonumber \\
+\frac{4F'ABM}{F}\delta_B &=& 0 ,
\end{eqnarray}

\begin{equation} \label{Nein7}
\frac{4F'ABM}{F}\gamma_B-\frac{4A}{B}\delta_B-\frac{4M}{B}\alpha_B=0 ,
\end{equation}
\begin{equation} \label{Nein8}
\frac{2F'BM^2}{F}\gamma_B-\frac{4M}{B}\delta_B=0 ,
\end{equation}
\begin{equation} \label{Nein9}
\frac{2F'BM^2}{F}\gamma_A-\frac{4M}{B}\delta_A=0 ,
\end{equation}
\begin{equation} \label{Nein10}
-BM^2U\alpha-AM^2U\beta-ABM^2U_\varphi\gamma-2ABMU\delta=0 .
\end{equation}

We can find two linearly independent vectors satisfying equations \eqref{Nein1} -- \eqref{Nein10}, which are
\begin{equation}
\label{noevec2}
\stackrel{\to }{N_3}=(0, 0, k, 0), \quad \stackrel{\to }{N_4}=(-2A, 0, 0, M), 
\end{equation}
where $K$ is a constant. Following the same procedure as in the Jordan frame, we find that
\begin{equation}
\label{fwu2}
F(\varphi)=k_1 e^{K\varphi},\\ \omega(\varphi)=k_2 e^{K\varphi},\\ U(\varphi)=U_0 ,
\end{equation}
with the corresponding constants of motion given respectively by

\begin{eqnarray} \label{sigma3}
\Sigma_3 &=& 2\sin(\theta)K\frac{\varphi_rAM^2\omega}{FB}+4\sin(\theta)K\frac{ABMM_rF'}{F} \nonumber \\
&& +2\sin(\theta)K\frac{A_rBM^2F'}{F}-3\sin(\theta)K\frac{\varphi_rAM^2F'^2}{F^2B} . \\
\Sigma_4 &=& 4\sin(\theta)\frac{AM-A_rM^2}{B} .  \label{sigma4}
\end{eqnarray}

We can eliminate $B$ using these two constants of motion. To do that, we firstly substitute the results \eqref{fwu2} into \eqref{sigma3}, and then use $M=r$ together with the constraint $2k_2=3k_1K^2$. Now we can integrate and find a solution for $A (r)$ given by
\begin{equation}
\label{A}
A(r)=\sqrt{\frac{\Sigma_3\Sigma_4}{k_1K}}\frac{1}{2r} .
\end{equation}
From \eqref{sigma4} and \eqref{A} it  follows immediately that
\begin{equation} \label{B1}
B(r)=4\sin(\theta)\sqrt{\frac{\Sigma_3}{\Sigma_4k_1K}} .
\end{equation}

This last expression may be used together with the Euler-Lagrange equation for $B(r)$ to integrate $\varphi(r)$. From the Euler-Lagrange equation for $B(r)$ we obtain
\begin{equation}
\label{B2}
	B=\sqrt{\frac{2AM^2F\omega\varphi_r^2+4AF^2M_r^2+8MF^2A_rM_r+3AM^2F'^2\varphi_r^2}{2AM^2F^2U-8AMFF'\varphi_rM_r-4M^2FF'\varphi_rA_r}}
\end{equation}
Equating \eqref{B1} and \eqref{B2} for $U=0$, integrating for $\varphi$ and renaming the constants we finally obtain
\begin{equation}
\label{phiR}
	\varphi(r)=c_4+c_5ln(\frac{1}{r})+c_6ln(r) .
\end{equation}


\section{Solutions for a cylindrically symmetric tetrad}
\subsection{Jordan frame}

In this section we are going to make similar calculations in the Jordan frame \eqref{LagJor} for a cylindrically symmetric spacetime. For this purpose we choose a static cylindrically symmetric spacetime \cite{prasanna:1975} and write the corresponding diagonal tetrad defined in cylindrical coordinate frame $(t, r, \phi, z)$ with general functional form given by
\begin{equation}\label{tet2}
{e}^{i}_{\phantom{i}\nu}=
\begin{bmatrix}
A(r) & 0 & 0 & 0\\
0 & C(r) & 0 & 0\\
0 & 0 & M(r) & 0\\
0 & 0 & 0 &  C(r)
\end{bmatrix} .
\end{equation}
Computing the point-like Lagrangian for the Jordan frame \eqref{f2} we obtain

\begin{eqnarray}
L_{JF} &=& -\varphi_r^2AM\omega-AC^2MU-\frac{2FC_rM_rA}{C} \nonumber \\
&& -2FA_rM_r-\frac{2FA_rC_rM}{C}.
\end{eqnarray}\

Applying the Noether Symmetry Approach as we did for the spherically symmetric tetrad we get the set of equations given by
\begin{equation} \label{cy1}
	AM\varOmega\gamma_M+\frac{FA}{C}\beta_\varphi+F\alpha_\varphi=0 ,
\end{equation}
\begin{equation} \label{cy2}
	AM\varOmega\gamma_C+\frac{FM}{C}\alpha_\varphi+\frac{FA}{C}\delta_\varphi=0 ,
\end{equation}
\begin{equation} \label{cy3}
	AM\varOmega\gamma_A+\frac{FM}{C}\beta_\varphi+F\delta_\varphi=0 , 
\end{equation}
\begin{equation} \label{cy4}
2AM\varOmega\gamma_\varphi+AM\varOmega_\varphi\gamma+M\varOmega\alpha+A\varOmega\delta=0 ,
\end{equation}
\begin{equation} \label{cy5}
F'\gamma+\frac{FM}{C}\beta_M+\frac{FA}{C}\beta_A+F\delta_M+F\alpha_A=0	,
\end{equation}
\begin{equation} \label{cy6}
\frac{F'M}{C}\gamma+\frac{FM}{C}\beta_C+\frac{FM}{C}\alpha_A-
\frac{FM}{C^2}\beta+\frac{FA}{C}\delta_A+\frac{F}{C}\delta+F\delta_C=0
\end{equation}
\begin{equation} \label{cy7}
\frac{FM}{C}\beta_A+F\delta_A=0 ,
\end{equation}
\begin{equation} \label{cy8}
\frac{FA}{C}\beta_M+F\alpha_M=0 ,
\end{equation}
\begin{equation} \label{cy9}
\frac{FM}{C}\alpha_C+\frac{FA}{C}\delta_C=0 ,
\end{equation}
\begin{equation} \label{cy10}
\frac{F'A}{C}\gamma+\frac{FM}{C}\alpha_M+\frac{FA}{C}\delta_M-
\frac{FA}{C}\beta_C+\frac{F}{C}\alpha-\frac{FA}{C^2}\beta+F\alpha_C=0
\end{equation}
\begin{equation} \label{cy11}
AC^2MV_\varphi\gamma+MC^2V\alpha+2ACMV\beta+AC^2V\delta=0 .
\end{equation}
We can immediately write down the solutions for $\alpha,\beta, \gamma,\delta$ as
\begin{align}
\stackrel{\to }{N_5}&=\begin{aligned}
&(A, 0, 0,-M),\end{aligned}\\
\stackrel{\to }{N_6}&=\begin{aligned}
(A,-2C, 0,M),\end{aligned}\\
\stackrel{\to }{N_7}&=(A, 0, \varphi, 0).
\end{align}
Considering these sets of values for $\alpha,\beta, \gamma,\delta$,  we find the functional forms of $F(\varphi), \omega(\varphi), U(\varphi)$ as
\begin{equation}
\label{fwuu}
F(\varphi)=\frac{k_3}{\varphi^2},\\ \omega(\varphi)=\frac{k_4}{\varphi^3} ,\\ U(\varphi)=\frac{k_5}{\varphi}.
\end{equation}

In order to integrate the tetrad potentials, we need constants of motion for the first two sets of values of $\alpha,\beta, \gamma,\delta$. We thus find
\begin{equation} \label{sigma5}
\Sigma_5=2F(M_rA-MA_r) ,
\end{equation}
\begin{equation} \label{sigma6}
\Sigma_6=2F(M_rA+MA_r-\frac{2MAC_r}{C}) .
\end{equation}
By using the relation $M=\frac{r}{A}$ \cite{prasanna:1975}, and assuming $\varphi(r)=ln(r)$, we  obtain
\begin{equation}
\label{Acyl}
	A(r)=K_0r^{\frac{\Sigma_5-2}{4k_3}}
\end{equation}
and 
\begin{equation}
	C(r)=K_1r^{1-\frac{\Sigma_6+2}{4k_3}} .
\end{equation}


\subsection{Einstein frame}

For \eqref{tet2} we obtain the Einstein frame Lagrangian \eqref{Lageins} as
\begin{eqnarray} \label{LagEin2}
L_{EF} &=& -A M C^2 U+\varphi_r^2(\frac{ A M \omega}{F }-\frac{3F_\varphi^2 A M}{F^2}) \nonumber \\
&& -\frac{2 M_r C_r A }{C}+\frac{2 \varphi_r M_r A C^2 F_\varphi}{F}+2 A_r M_r \nonumber \\
&& +\frac{2 \varphi_r C_r ACM F_\varphi }{F}-\frac{2C_r A_r M}{C}+\frac{2A_r \varphi_r C^2 M F_\varphi}{F} .
\end{eqnarray} \\
We then apply the Noether Symmetry Approach as we did for Jordan frame and find the equations given by

\begin{eqnarray}  \label{Nein22}
-(\frac{2AM\omega}{F}+\frac{3F'^2AM}{F^2})\gamma_A+\frac{2F'CM^2}{F}\gamma_\varphi && \nonumber \\
+(\frac{2F"MC^2}{F}-\frac{2F'^2MC^2}{F^2})\gamma+\frac{2F'MC^2}{F}\alpha_A && \nonumber \\
+\frac{2F'ACM}{F}\beta_A+\frac{4F'CM}{F}\beta && \nonumber \\
-\frac{2M}{C}\beta_\varphi+\frac{2AC^2F'}{F}\delta_A-2\delta_\varphi+2\frac{F'C^2}{F}\delta &=& 0 ,
\end{eqnarray}

\begin{equation} \label{Nein44}
\frac{2F'AC^2}{F}\gamma_M-\frac{2A}{C}\beta_M-2\alpha_M=0 ,
\end{equation}

\begin{eqnarray} \label{Nein55}
\frac{2F'MAC}{F}\gamma_M+\frac{2F'AC^2}{F}\gamma_C-\frac{2M}{C}\alpha_M && \nonumber \\
-\frac{2A}{C}\beta_C -\frac{2A}{C}\delta_M-\frac{2\alpha}{C}+\frac{2A\beta}{C^2}-2\alpha_C &=& 0 ,
\end{eqnarray}

\begin{equation} \label{Nein66}
\frac{2F'M^2C}{F}\gamma_M+\frac{2F'AC^2}{F}\gamma_A-\frac{2M}{C}\beta_M -\frac{2A}{C}\beta_A -2\delta_M-2\alpha_A = 0 ,
\end{equation}

\begin{eqnarray}  \label{Nein77}
-(\frac{2AM\omega}{F}+\frac{3F'^2AM}{F^2})\gamma_C+\frac{2F'CMA}{F}\gamma_\varphi && \nonumber \\
+(\frac{2F"MCA}{F}-\frac{2F'^2MCA}{F^2})\gamma+\frac{2F'MC^2}{F}\alpha_C && \nonumber \\
+\frac{2F'ACM}{F}\beta_C+\frac{2F'CM}{F}\alpha-\frac{2M}{C}\alpha_\varphi&&\nonumber\\
+\frac{2AMF'}{F}\beta+2\frac{F'AC^2}{F}\delta_C+2\frac{F'AC}{F}\delta-\frac{2A}{C}\delta_\varphi &=& 0 ,
\end{eqnarray}

\begin{eqnarray} \label{Nein33}
-(\frac{2AM\omega}{F}+\frac{3F'^2AM}{F^2})\gamma_\varphi-\frac{AM\omega_\varphi}{FB}\gamma && \nonumber \\
+\frac{2F'AM\omega}{F^2}\gamma-\frac{3F'F"AM}{F^2}\gamma+\frac{3F'^3AM}{F^3}\gamma && \nonumber \\
-(\frac{M\omega}{F}+\frac{3F'^2M}{2F^2})\alpha+\frac{2F'MC^2}{F}\alpha_\varphi && \nonumber \\
-(\frac{A\omega}{F}+\frac{3F'^2A}{2F^2})\delta 
+\frac{2F'AC^2}{F}\delta_\varphi+\frac{2F'ACM}{F}\beta_\varphi &=& 0 ,
\end{eqnarray}

\begin{equation}\label{Nein11}
	\frac{2F'ACM}{F}\gamma_C-\frac{2A}{C}\delta_C = 0 ,
\end{equation}

\begin{eqnarray}\label{Nein88}
	\frac{2F'CM^2}{F}\gamma_C+\frac{2F'ACM}{F}\gamma_A-\frac{2M}{C}\beta_C && \nonumber \\
	-\frac{2M}{C}\alpha_A+\frac{2M}{C^2}\beta-\frac{2A}{C}\delta_A-\frac{2}{C}\delta-2\delta_C &=& 0 ,
\end{eqnarray}

\begin{equation}
	\frac{2F'CM^2}{F}\gamma_A-\frac{2M}{C}\beta_A-2\delta_A = 0 ,
\end{equation}

\begin{equation}
AC^2MU'\gamma+MC^2U\alpha+2ACMU\beta+AC^2U\delta = 0 ,
\end{equation}

\begin{eqnarray}  \label{Nein99}
-\frac{3F'^2MA}{F^2})\gamma_M+\frac{2F'AC^2}{F}\gamma_\varphi +\frac{2F'C^2M}{F}\alpha_M && \nonumber \\
+\frac{2F'ACM}{F}\beta_M+\frac{2AC^2F'}{F}\delta_M+2\frac{F'C^2}{F}\alpha && \nonumber \\
+4\frac{F'CA}{F}\beta-\frac{2A}{C}\beta_\varphi-2\alpha_\varphi && \nonumber \\
+(\frac{2F"AC^2}{F} -\frac{2F'^2AC^2}{F^2})\gamma +\frac{2\omega AM}{F}\gamma_M&=& 0 ,
\end{eqnarray} \\

Two solutions are immediately found for \eqref{Nein22}-\eqref{Nein99} as
\begin{align}
\stackrel{\to }{N_8}&=
(A, 0, 0,-M),\\
\stackrel{\to }{N_9}&=
(0,0, k,0) ,
\end{align}
where $k$ is a constant. Assuming $M=\frac{r}{A}$,  we find the corresponding  constants of motion as
\begin{equation}
	\Sigma_8=2MA_r-2AM_r ,
\end{equation}
\begin{eqnarray}
	\Sigma_9=k(-\varphi_r(\frac{2MA\omega}{F}+\frac{3MAF'^2}{F^2})+M_r\frac{2AC^2F'}{F}&& \nonumber \\ 
	+ C_r\frac{2ACMF'}{F} +A_r\frac{2MC^2F'}{F} ) .
\end{eqnarray}
From these two solutions,  we find $F(\varphi), \omega(\varphi),  U(\varphi)$ as
\begin{equation}
	F(\varphi)=k_6 e^{K\varphi},\\ \omega(\varphi)=k_7 e^{-2K\varphi},\\ U(\varphi)=U_0 .
\end{equation}
Using the assumptions we made before, $M=\frac{r}{A}$ and  $\varphi(r)=ln(r)$, we obtain $A(r)$ as
\begin{equation}
A(r) = c_7r^{\frac{\Sigma_8+2}{4}} ,
\end{equation}
and after rearranging the constants, we obtain $C(r)$ as
\begin{equation}
	C(r) = \sqrt{-\frac{c_8}{r^{3K}}-\frac{c_9}{r^{2}}+c_{10}}
\end{equation}

\section{Conclusions}
In this study we used the method proposed in  \cite{capAxial} in order to find exact solutions for  $f(T)$ gravity  for both spherical and cylindrically symmetric. However, due to the absence of Lorentz invariance, the approach is valid if a tetrad is fixed and the related metric is derived.

The method is  based on searching for Noether symmetries of the Lagrangians in  the Jordan and the Einstein frames. Similar to \cite{capAxial},  we used constants of motion coming from  the Noether symmetries. The main difference was the use of multiple constants of motion in order to eliminate some of the equations to obtain the  full integration. We first used Jordan frame Lagrangian for a spherically symmetric tetrad and, by  two constants of motion,  we eliminated $B(r)$ and integrated $A(r)$. After that, by using the assumption $\varphi(r)=ln(r)$,  we calculated $B(r)$. When we switched to the Einstein frame, we faced a more complicated structure. Nevertheless, besides $A(r)$ and $B(r)$, the Einstein frame gave us the opportunity to integrate $\varphi(r)$. Lastly, we obtained the Jordan and Einstein frame Lagrangians for a general cylindrically symmetric tetrad. Following the same procedure as we did for spherically symmetric case, we calculated the tetrad potentials $A(r)$ and $C(r)$ with the help of two constants of motion with three generating vectors in the Jordan frame and two constants of motion with two generating vectors in the Einstein frame.

As final remark, it is worth stressing that the possibility to achieve solutions in both frames by the existence of Noether symmetries can help in interpreting the physical meaning of the Jordan and the Einstein frame. For a discussion of this topic, see \cite{tsamparlis}.

\begin{acknowledgements}
A.N.N. would like to thank Prof. Talat Sayga\c{c} for his support during this work.
 S.C. acknowledges INFN Sez. di Napoli ({\it Iniziative Specifiche} QGSKY and MOONLIGHT2) for support. This article is also based upon work from COST action CA15117 (CANTATA), supported by COST (European Cooperation in Science and Technology). 
 \end{acknowledgements}

\end{document}